\documentclass[letterpaper]{JHEP3}
\usepackage{epsfig,verbatim}
\usepackage{subfigure}
\usepackage{amsmath, amssymb, graphics}
\usepackage{comment}

\newcommand{\mathsym}[1]{{}}

\newcommand{\eref}[1]{(\ref{#1})}

\renewcommand\({\left(}
\renewcommand\){\right)}
\renewcommand\[{\left[}
\renewcommand\]{\right]}

\newcommand{\e}{{\rm e}}

\newcommand\eps{\epsilon}

\def\ba{\begin{eqnarray}}
\def\ea{\end{eqnarray}}
\def\be{\begin{equation}}
\def\ee{\end{equation}}

\def\M{\mathcal{M}}

\def\nn{\nonumber}
\def\({\left(}
\def\){\right)}

\def\eref#1{(\ref{#1})}

\newcommand{\roughly}[1]{\mathrel{\raise.3ex\hbox{$#1$\kern-0.85em
\lower1ex\hbox{$\sim$}}}}

\title{Sgoldstino inflation}

\author{
Ana Ach\'ucarro$^{a,b}$, Sander Mooij$^c$, Pablo Ortiz$^{a,c}$ and Marieke Postma$^c$  \\ 
\\
$^a$ Instituut-Lorentz Theoretical Physics, Universiteit Leiden, Niels Bohrweg 2, 2333 CA Leiden, The Netherlands\\
$^b$ Department of Theoretical Physics, University of the Basque Country, 48080 Bilbao, Spain \\
$^c$ Nikhef, Science Park 105, 1098 XG Amsterdam, The Netherlands 
 
}

\date{}

\abstract {We discuss the possibility that inflation is driven by the
  sgoldstino, the superpartner of the goldstino. Unlike in generic
  supergravity scenarios, the sgoldstino decouples from all other
  fields in the theory, which allows for a simple and robust
  inflationary model.  We argue that the two-field model given by this
  single complex scalar correctly captures the full multifield
  inflationary phenomenology. On the other hand, the assumption of
  stability, along the entire inflationary trajectory, of the
  supersymmetry--preserving sector that is integrated out leads to
  supplementary constraints on the parent supergravity.  We
  investigate small field, large field and hybrid sgoldstino inflation
  scenarios and provide some working examples. They are subject to the
  usual fine-tuning issues that are common to all su\-per\-gra\-vi\-ty models
  of inflation. We comment on some other recently proposed sgoldstino
  inflation models.}

\preprint{NIKHEF 2012-003}

\begin{document}


\section{Introduction}

Scalar fields are abundant in supersymmetric theories.  They all
couple to each other with at least gravitational strength
interactions.  Planck suppressed couplings are generically unimportant
when describing processes at low energies, but such a decoupling does
not work for inflation.  This can be most easily inferred from the
slow roll parameters, which get contributions from dimension five and
six operators that are unsuppressed.  Describing inflation in a
generic supergravity model is thus a challenging task, as generically
the scalar field dynamics pose a complicated multifield problem.

There are good reasons why a single--field description is
desirable. In line with Ockham's razor, it is the simplest model that
fits the data. Multifield slow--roll inflation with several (real)
light fields has been studied for over a decade
\cite{Gordon:2000hv,GrootNibbelink:2000vx,GrootNibbelink:2001qt,Wands:2002bn}
(see \cite{Malik, Peterson} and references therein),
and is very constrained by the observations, in particular through the
tight limits on isocurvature modes and non-gaussianity \cite{WMAP7}.
On the other hand, however, it is technically challenging to obtain
single-field behavior in a full multi-field set-up.

If the inflaton is the only light field in a multifield parent theory,
integrating out the heavy fields should yield an effective
single--field description that is accurate up to terms
$\mathcal{O}(\partial^2/M^2)$, with $M$ the mass of the heavy
field. Naively, if there is slow roll and a large mass hierarchy, one
would assume such terms can be ignored, but this expectation is
premature\footnote{The caveats are due to other mass scales introduced
  by the changing background, whether in flat space or during
  slow-roll inflation \cite{Achucarro:2010jv}. This makes the details
  of decoupling during inflation different from particle
  phenomenology, where the effective field theory expansion is around
  a particular vacuum.}. In particular, if there are turns in the
inflationary trajectory, derivative interactions between the inflaton
and the heavy fields can become transiently strongly coupled. These
lead to features and non-gaussianity in the spectrum of primordial
perturbations that would not be inferred from the naive effective
field theory (EFT). If the heavy fields remain sufficiently massive,
the turns result in a reduced speed of sound for the adiabatic
perturbations but are otherwise completely consistent with slow-roll
\cite{Achucarro:2010jv,Achucarro:2010da,Cespedes,Shiu,Baumann}. Careful
integration of the heavy fields recovers the general low energy
effective field theory of inflation including a variable speed of
sound for the adiabatic perturbations
\cite{Cheung:2007st,Weinberg:2008hq,Tolley,Garriga,cliff1} (see
\cite{Achucarro:2012sm} for a detailed discussion).

These interactions are unavoidable whenever the potential ``valley''
provided by the multifield potential deviates from a geodesic of the
multifield sigma model metric. A corollary from the point of view of
inflationary model building is that, when it comes to precision
cosmology, the field space geometry of the ``spectator'' heavy fields
(that are supposed to sit at their adiabatic minima during slow--roll
inflation) is as important as their masses and couplings inferred from
the potential alone. 

Among the many scalars in a supersymmetric theory, the sgoldstino
field stands out. The sgoldstino is the scalar partner of the
goldstino, and belongs to the chiral superfield whose non-zero F-term
breaks supersymmetry\footnote{We will not consider D-term breaking in
  this work.}. It has the special property \cite{
Binetruy:2004hh, kepathesis,Achucarro:2010jv} that it decouples from all
other fields in the theory\footnote{More precisely, setting all
  other superfields at the minimum of their potential is a consistent
  truncation from the N=1 sugra multi--field parent theory to an
  effective N=1 sugra with a single chiral superfield, the sgoldstino.
  In particular, the (real, two-dimensional) sgoldstino plane is a
  geodesically generated surface of the K\"ahler metric in the parent
  theory, so there are no derivative interactions with the truncated
  heavy fields: all turns in the inflationary trajectory are entirely
  confined to the sgoldstino plane. The effects of the fluctuations of
  the heavy fields are suppressed by their mass squared exactly as one
  would expect from an EFT expansion.}.
This makes the sgoldstino an ideal inflaton candidate, for it allows
for a description of inflation in terms of a single complex
field. From the point of view of inflationary modeling this is still
multifield inflation (with two real fields), but this two--field model
is not a toy model, it really is the correct effective description for
the full multifield system.

If inflation is effectively driven by a single real scalar field, the
inflaton, this can be identified with a suitable linear combination of
the real and imaginary parts of the sgoldstino field. Meanwhile, the
orthogonal combination is to remain stabilized at a local minimum of
the potential during inflation. The effect of turns in the trajectory
on the spectrum of primordial perturbations have to be taken into
account when comparing to observations, but at least they can be
calculated from the two-field model (see
\cite{Chen,Achucarro:2010da,PetersonTegmark,Cremonini, Avgoustidis} for
recent discussions and references).

Needless to say, this does not mean that all other scalars in the
theory (from the susy-preserving superfields) can be completely
neglected, as they have to be stabilized in a minimum of the potential
during inflation.  Even though the sgoldstino decouples from these
fields, vice versa this is not true: the masses and couplings of all
other fields generically depend on the field value of the sgoldstino
field.  As during inflation the sgoldstino evolves along its
inflationary trajectory, the masses of the scalars change. If the
inflaton is the sgoldstino, they will remain at the critical points,
but they may become light or even tachyonic, triggering a
waterfall-type exit from inflation that is not seen in the two-field
model.  Although it is still a complicated task to determine the
minimum of the multifield potential along the inflationary trajectory,
it is much simpler than the full multifield {\it dynamics} needed for
a generic, non-sgoldstino, inflation model.

The potential energy density driving inflation breaks susy
spontaneously \cite{Dine:1995kz, Dine:1995uk}.  This source of susy
breaking in the inflaton sector is always present during inflation,
and is in principle independent of the source of vacuum susy breaking.
For sgoldstino inflation there are two possibilities.  First, the same
superfield that drives inflation is also responsible for low energy
susy breaking\footnote{This possibility has been recently discussed in
  \cite{minimal,minimal1,minimal2} but as we will show it is difficult
  to implement in practice.}. This would be the ideal situation. Not
only does inflation decouple from all other fields in the theory, it
also links the scale of inflation to the scale of susy breaking.  The
second possibility is that the two sources of susy breaking are due to
different fields. Both sources may be operative during inflation, or
alternatively, it may be that only after inflation has ended, a phase
transition takes place generating our present-day susy breaking.  In
both cases the present day sgoldstino field is not the sgoldstino
during inflation.

If several sources of susy breaking are present during inflation, the
inflaton can only be approximately identified with the sgoldstino, and
only if the vacuum susy breaking scale is much below the inflationary
scale.  Care should be taken in this case because, as argued in
\cite{Hardeman}, only if the lightest mass in the hidden sector
responsible for vacuum susy breaking is much larger than the inflaton
mass and if the inflaton mass is much larger than the scale of hidden
sector susy breaking, is the effect of the hidden sector on the
slow-roll dynamics of the inflaton negligible. This is far from
trivial; for example, it has been proven extremely hard to combine a
susy breaking moduli stabilization and inflation in a consistent way
\cite{Kallosh:2004yh}, even in a fine-tuned setting
\cite{DavisPostma1}.

The decoupling of the sgoldstino from the other fields fits in with
recent work on how to incorporate different fields, or sets of fields,
in a sugra set-up minimizing the coupling between them
\cite{Binetruy:2004hh,deAlwis1,deAlwis2,Kepa1,Kepa2,Kepa3,serone1,serone2,gallego,Brizi,Hardeman,Achucarro:2011yc}.
Quite commonly different sectors --- e.g. the fields and couplings
responsible for susy breaking, for inflation, for moduli
stabilization, or making up the standard model --- are combined by
simply adding their respective K\"ahler- and superpotentials. However,
following this procedure one cannot completely decouple these
sectors. Even if the K\"ahler and superpotential do not contain direct
interaction terms between fields in different sectors, the resulting
scalar potential does. There are always at least Planck suppressed
interactions between the fields, and generically the mass matrix is
not block dia\-go\-nal in the different sectors.  This complicates the
analysis of the full model enormously.  Sectors are affected by the
presence of others, and although they work in isolation, they may no
longer do so in the full set-up.  Moreover, heavy fields generically
cannot be integrated out in a consistent supersymmetric
way.\footnote{Here, once again, approximations that are justified for
  phenomenology applications where the background is static
  \cite{Brizi} fail during inflation
  \cite{deAlwis1,deAlwis2,Kepa3,Achucarro:2010jv}}

The cross-couplings between sectors can be minimized if instead of
adding K\"ahler and superpotentials, one adds the K\"ahler invariant
functions $G = K + \ln|W|^2$ for the two sectors \cite{Binetruy,Kepa1}.  This
approach allows to integrate out fields in a susy preserving
way \cite{Binetruy:2004hh}. In Ref. \cite{Kepa1} the addition of sugra
functions was used to couple a susy breaking moduli sector (fields
$X^i$) to a susy preserving sector, for example the standard model
(fields $z_i$):
\be
G^{\rm tot}(X^i,\bar{X^i},z_i,\bar z_i) = g(X^i,\bar{X^i}) +
G^{\rm other}(z_i,\bar z_i). 
\label{G_ana} 
\ee 
In this article we want to use the same idea to couple a susy breaking
inflationary sector (fields $X^i$) to a susy preserving sector
($z_k$)\footnote{In \cite{DavisPostma2} the separable form
  \eref{G_ana} was used to combine hybrid inflation with a susy
  breaking moduli sector in a successful way.  In this set-up the
  inflaton is not the goldstino.}. For simplicity we restrict to
effectively single field inflation, and models with a single inflaton
field $X$.  As susy is broken during inflation, the inflaton is then
the sgoldstino. As it turns out, the ansatz \eref{G_ana} is actually
too strict.  We can allow for explicit couplings between the inflaton
and the other fields, of the form
\ba
G(X,\bar X,z_k ,\bar z_k) = g(X,\bar X) + \frac12 \sum_{i \geq j} &\bigg[&
(z_i-(z_i)_0) (z_j-(z_j)_0)  f^{(ij)}(X,\bar X,z_k ,\bar z_k) 
 \nn \\
+
&&(z_i-(z_i)_0) (\bar z_j-(\bar z_j)_0)  h^{(ij)}(X,\bar X,z_k ,\bar z_k)+{\rm h.c.
  }\bigg]
\label{G_intro}
 \ea
 with $f,h$ arbitrary functions of its arguments. As we will show,
 this is the most general ansatz consistent with X being the
 sgoldstino.  The explicit $X$-dependence in the second term does not
 spoil the decoupling of the inflaton field, the mass matrix remains
 block diagonal in the two sectors, as long as the fields $z_i$ sit at
 the susy critical point $(z_i)_0$ during inflation.  As we will show,
 during sgoldstino inflation the K\"ahler function $G$ is well
 defined, maybe except from isolated points in field space.

 Single field inflation can be divided into three main classes: large
 field, small field and hybrid inflation. We discuss whether and how
 sgoldstino inflation might work in these three regimes. Any sugra
 model of inflation has to address the $\eta$-problem; this puts
 bounds on the K\"ahler geometry \cite{Covi1,Covi2,BenDayan}.

Large field sgoldstino inflation does not work, at least not for potentials that grow at most polynomial.

Hybrid inflation provides the most natural embedding for sgoldstino
inflation.  Indeed, usual F-term hybrid inflation is an example of
having a sgoldstino inflaton.  In this set-up susy is restored in the
vacuum, and there is no relation with low energy susy breaking.  More
complicated waterfall regimes may be devised, such that susy is broken
in the minimum after inflation. However, such an analysis is
multifield, and complicated multifield dynamics enters via the back
door again.

Small field inflation offers the best possibility to link inflation to
susy breaking.  Naively all that is needed is finding and tuning a
saddle or maximum in a single field potential with a susy breaking
Minkowski minimum.  We only find inflection points suitable for
inflation rather than a maximum or saddle point. Inflection point
inflation yields \cite{brax, Linde:2007jn} a low spectral index $n_s \leq 0.92-0.93$ (for $N =
50-60$ efolds), which is on the verge of being ruled out by the CMB
data \cite{WMAP7}.  Interestingly enough, models in which susy is
broken after inflation are much easier to embed in a multi-field
set-up than models with a susy preserving vacuum.  Finally, we comment
on recent claims in the literature for small field sgoldstino
inflation \cite{minimal1,minimal2,dine} with no or very little
fine-tuning.  We will explain why these models cannot work.


\section{Decoupling of the sgoldstino}
\label{s:decoupling}

In this section we will show the decoupling of the sgoldstino field
explicitly. In the first subsection we derive the mass matrix, which
is block diagonal along the sgoldstino inflation trajectory. We will
argue in subsection \ref{s:G} that the K\"ahler function for a
dynamical sgoldstino field can always be put in the form
\eref{G_intro}.  In subsection \ref{s:traject} we show that this
sgoldstino trajectory is independent of the field values of all the
other fields. Vice versa that is not the case: the dynamics of the
non-sgoldstino fields does depend on the sgoldstino field.  Care must
be taken so that these fields remain stabilized along the full
inflationary trajectory.  Finally, in subsection \ref{s:adding} we
discuss the special limit of separable K\"ahler functions
\eref{G_ana}, in which the results of \cite{Kepa1} are retrieved.


\subsection{Mass matrix}
\label{s:mass}

We start with the general formula for the mass matrix, then specialize
to sgoldstino inflation.  The scalar potential can be expressed solely
in terms of the K\"ahler function\footnote{This procedure is ill
  defined for $W=0$. To cure this, one can use the variable $\phi
  \equiv \e^G$ instead, which remains well defined \cite{Barbieri}.
  However, in the next section we show that $W=0$ at most in isolated
  points in field space.} $G = K + \ln |W|^2$:
\be
V_F = \e^G [G_I G^{I\bar J} G_{\bar J} -3],
\label{VF}
\ee
with $I,J$ running over all fields $\Phi_I$.  We will be working in
Planck units $M=1$ throughout this work. The fields span the K\"ahler
manifold with complex metric $G_{I \bar J}$.  The inverse metric
$G^{I\bar J}$ is such that $G_{I\bar J}\, G^{K\bar J}=\delta_I^K$ and
$G_{I\bar J}\, G^{I\bar K}=\delta_{\bar J}^{\bar K}$.  The only
non-zero connection is $\Gamma_{IJ}^K = G_{IJ\bar P}G^{\bar P K}$ and
its complex conjugate. The non-zero components of the Riemann tensor
are $R_{I\bar J K \bar L} =G_{S\bar L} \partial_{\bar J} \Gamma^{
  S}_{IK}$ and permutations thereof.

The mass matrix is
\be
\M = \( 
\begin{array}{cc} 
M^I_J & M^I_{\bar J} \\
M^{\bar I}_J & M^{\bar I}_{\bar J}
\end{array} \), 
\qquad 
M^I_J = G^{I  \bar K} \nabla_{\bar K} \nabla_J V,
\quad
M^I_{\bar J}= G^{I \bar K} \nabla_{\bar K} \nabla_{\bar J} V,
\label{M}
\ee
with $\nabla_K v_L = \partial_K v_L - \Gamma_{KL}^M v_M$ the covariant
derivative of some vector $v_L$. The eigen\-va\-lues and eigenvectors of
the mass matrix correspond to the $m^2$--values and mass
eigenstates respectively.  The first derivative of the potential is
\be
V_K = G_K V + \e^G[G^I \nabla_K G_I + G_K ]
\label{dV}
\ee
where we used metric compatibility $\nabla_K G_{I\bar J} = 0$,
$\nabla_K G^I = \delta^I_K$ and introduced the notation $V_K
= \partial_K V$, $G^I = G^{I\bar J} G_{\bar J}$. Stationarity is not
assumed, as the inflaton field is displaced from its minimum during
inflation.  The second derivatives of the potential are
\ba
\nabla_{\bar L} \nabla_K V &=& (G_{K \bar L} -G_K G_{\bar L} )V +2 G_{(K}V_{\bar L)}
+\e^G[ G^{I\bar J}(\nabla_{\bar L} G_{\bar J})(\nabla_K G_I) 
- R_{I\bar J K \bar L}G^I G^{\bar J}
+G_{K\bar L}],
\nn \\
\nabla_{L} \nabla_K V &=& (\nabla_{(L} G_{K)} -G_{(K} G_{L)} )V +2 G_{(K}V_{L)}
+\e^G[ 2 \nabla_{(K}G_{L)} +G^I \nabla_{(L} \nabla_{K)} G_I],
\label{ddV}
\ea
where round brackets denote symmetrization. We used that
$[\nabla_{\bar L},\nabla_K]G_I=\nabla_{\bar L} \nabla_K G_I = -R_{K
  \bar L I \bar J}G^{\bar J}$.  Apart from the terms proportional to
$V_K$, which are absent for stationary situations, these equations are
the same as (2.6, 2.7) of Ref.~\cite{Covi}.

Now consider F-term breaking of susy, signaled by a non-zero $G_X \neq
0$.  Here $X$ is the scalar component of the chiral superfield which
also contains the goldstino. Note that one can always make a field
redefinition such that only one linear combination of fields breaks
susy. All other fields in the theory, denoted by $z_i$ (indexed by
lower case latin letters), do not break susy.  Hence, we split the
fields in $\Phi_I= \{X,z_i\}$, with
\be
G_X|_{z_0} \neq 0, \qquad G_{i}|_{z_0} = 0
\label{condition}
\ee
at some point in field space $z_0 = \{(z_1)_0, (z_2)_0, ...\}$, the so-called susy critical point. 

We are interested in a cosmological situation, in which $X(t)$ is
the inflaton rolling along some trajectory with $V_X \neq 0$. While
the inflaton rolls in the $X$-direction, we want all orthogonal fields
$z_i$ to remain extremized at $z_0$. To that end we demand that
\be
\left(\partial_{X}\right)^m\left(\partial_{\bar X}\right)^n G_i|_{z_0}
=0 ,
\qquad \forall m,n \in \mathbb N.
\label{condition2}
\ee
Indeed, from \eref{dV}, we then have that
\be
V_i|_{z_0} = G_i V + \e^G[G^P \nabla_{i} G_P + G_{i}]
= \e^G G^X \nabla_{i} G_X =0.
\ee
For notational convenience we dropped the $|_{z_0}$ on the right hand
side, but the reader should keep in mind that all expressions should
be evaluated at $z = z_0$.  Note that $i$ labels the $z_i$ fields, and
capital letters label $\Phi_I$ (i.e. also running over $X$). In
the first step we used \eref{condition}, in the second $\nabla_{i}
G_X|_{z_0} =0$, which is a consequence of \eref{condition2}.

Thus $z_i = (z_i)_0$ is an extremum of the potential.  To see whether
this is a maximum, minimum or saddle point, we must calculate the
eigenvalues of the mass matrix, which need to be positive definite for
a stable minimum.  This analysis is simplified because \eref{condition} assures
the mass matrix is in block diagonal form, i.e. $M^X_{i}|_{z_0} =
M^{\bar X}_{i}|_{z_0} =0$. To prove this last statement, it is enough
to show the block diagonal form of the second covariant derivatives,
as it follows from \eref{condition2} that the field metric $G_{I\bar
  J}|_{z_0}$ is block diagonal as well.  The first equation in
\eref{ddV} gives for mixed indices
\ba
\nabla_{\bar i} \nabla_X V |_{z_0}&=& 
(G_{X \bar i} -G_X G_{\bar i} )V +2 G_{(X}V_{\bar i)}
+\e^G[ G^{K \bar L}(\nabla_{\bar i} G_{\bar L})(\nabla_X G_K) 
- R_{K \bar L X \bar i}G^K G^{\bar L}
+G_{X\bar i}]\nn \\
&=& -\e^G G^X G^{\bar X}R_{X \bar X X \bar i} =0.
\ea
In the first step we used (\ref{condition}, \ref{condition2}) and that
$\nabla_i G_X|_{z_0} =\nabla_X G_i|_{z_0} =0$; in the second step that
$R_{X \bar X X \bar i}|_{z_0} = G_{j\bar i} \partial_{\bar X}
\Gamma^{j}_{ X X} = 0$ as well, which also follows from
\eref{condition2}. The second equation in \eref{ddV} likewise
vanishes for mixed indices:
\be
\nabla_{i} \nabla_X V|_{z_0} = (\nabla_{(i} G_{X)} -G_{(X} G_{i)} )V +2 G_{(X}V_{i)}
+\e^G[ 2 \nabla_{(X}G_{i)} +G^K \nabla_{(i} \nabla_{X)} G_K]=0.
\ee
%

\subsection{K\"ahler invariant function for sgoldstino inflation}
\label{s:G}

Let us quickly comment on our use of the K\"ahler invariant function $G = K +
\ln|W|^2$, rather than expressing results in terms of the K\"ahler
potential and superpotential.  The potential danger in using $G$ is that it becomes undefined when $W=0$. However, it is easy to show that for sgoldstino
inflation we nowhere have $W=0$, except maybe for isolated points in field
space. Therefore the K\"ahler function $G$ is well defined.  To illustrate this, consider a
theory with two chiral fields --- the extension to many fields is
straightforward --- with a superpotential $W=W(X,Z)$.  For sgoldstino
inflation, with $X$ the goldstino superfield, we have
\be
D_X W|_{\{X(t),Z_0\}} \neq0, \qquad D_Z W|_{\{X(t),Z_0\}}=0,
\label{demands}
\ee
with $D_XW = K_X W + W_X$ the K\"ahler covariant derivative.  Setting
$W=0$ along the {\it whole} trajectory implies \be W|_{\{X(t),Z_0\}}=0
\quad \Rightarrow \quad W_X|_{\{X(t),Z_0\}}=0 \quad \Rightarrow \quad
D_X W|_{\{X(t),Z_0\}}=0 \ee in contradiction with \eref{demands}.
Therefore the superpotential can only vanish for sgoldstino inflation
at accidental zeroes at isolated points in field space (possibly on
the trajectory, but this does not change our conclusions).

As a side remark, note that when the inflaton is identified with the $Z$
field rather than the sgoldstino field $X$,  as for example in the models of
Ref. \cite{rube}, it is possible to have $W=0$, $D_X W|_{\{X_0,Z(t)\}} \neq0$ and $D_Z W|_{\{X_0,Z(t)\}}=0$
along the whole trajectory $\{X_0,Z(t)\}$.  In this case the K\"ahler invariant
function is not well defined, and a description in terms of $K$ and
$W$ is needed.

Expanding the K\"ahler function around the susy critical point $z^i =
z_0^i$, the most general form for sgoldstino inflation --- satisfying
\eref{condition} and \eref{condition2} --- can be written in the form
\ba
G(X,\bar X,z_k ,\bar z_k) = g(X,\bar X) + \frac12 \sum_{i \geq j} &\bigg[&
(z_i-(z_i)_0) (z_j-(z_j)_0)  f^{(ij)}(X,\bar X,z_k ,\bar z_k) 
 \nn\\
+
&&(z_i-(z_i)_0) (\bar z_j-(\bar z_j)_0)  h^{(ij)}(X,\bar X,z_k ,\bar z_k)+{\rm h.c.
  }\bigg]\nn\\ \label{G} 
  \ea
  with $f,h,g$ arbitrary functions of its arguments.

\subsection{Inflationary trajectory}
\label{s:traject}

We have seen in subsection \ref{s:mass} that along the inflationary trajectory
all non-sgoldstino fields are extremized at $z^i = z_0^i$.  Since the
mass matrix is block diagonal, we can determine the stability of the
$z_i$ extremum from the sub-block of $\M$ with $z_i$ indices.  It can easily be shown that the inflaton trajectory itself is independent of
the field values of the other fields. Indeed, the potential along the
inflationary trajectory only depends on the function $g(X,\bar X)$ in
\eref{G}, and is thus independent of the field values of all other
fields.  The height $V_0 \equiv V|_{z_0}$, slope and second
derivatives of the inflaton potential are given by 
(\ref{VF},~\ref{dV},~\ref{ddV}) with $I,J$ only running over $X$ and
$G \to g$. For example we have
\ba
V_0 & =&\e^g\[g_X g^{X \bar X} g_{\bar X} -3\], \\
V_X|_{z_0} &=& g_X V_0 + \e^g\[g^X \nabla_X g_X + g_X\].
\ea

In contrast, the mass matrix along the orthogonal directions does
depend on the inflaton field value.  We find
\ba
M^{i}_{j}|_{z_0} &=& G^{i \bar k} \nabla_{\bar k} \nabla_{j} V\nn \\
&=&  G^{i \bar k} 
\[ G_{{j} \bar k} V_0
+\e^G[ G^{l \bar m}(\nabla_{\bar k} G_{\bar m})(\nabla_{j} G_l) 
- R_{X \bar X j \bar k}G^X G^{\bar X}
+G_{j\bar k}]\] \nn \\
&=& \e^g\[ \delta^i_j(b+1) + x^i_{\bar m} x^{\bar m}_j + w^i_j \],
\ea
and
\ba
M^{\bar i}_{j}|_{z_0} 
 &=& G^{\bar i k} \nabla_{k} \nabla_{j} V  \nn\\
&=&  G^{\bar i k} \[
\nabla_{(k} G_{j)} V_0 
+\e^G[ 2 \nabla_{(j}G_{k)} +G^X \nabla_{(k} \nabla_{j)} G_X] \]
\nn \\
&=& \e^g\[x^{\bar i}_j (b+2)+ y^{\bar i}_j \].
\ea
Here we introduced the notation
\ba
b & = & V_0e^{-g} = g_{_X} g^{_X} -3 
\label{b} \\
x^{\bar i}_j &= & 
G^{\bar i k}\nabla_k G_m = G^{\bar i k}\nabla_m G_k
\label{x} \\
w^i_j & =& -G^{i\bar k}G^X G^{\bar X} R_{X \bar X j \bar k}
\label{w} \\
y^{\bar i}_{j} &=& G^{\bar i k}G^{X} \nabla_{(k} \nabla_{j)} G_{X}.
\label{y}
\ea
Note that $b = V_0/m_{3/2}^2$ gives the height of the potential in
units of the gravitino mass. During slow-roll this is approximately $b
\sim 3H^2/m_{3/2}^2$.

The functions $b,x,y,w$ can be expressed in terms of the functions
$f,g,h$ appearing in the K\"ahler function \eref{G}. In general, the
constraint that the squared masses should be positive is complicated,
but there are two situations in which it simplifies considerably. The
first one, discussed in the next section, is if the K\"ahler invariant
function is separable \cite{Kepa1,Kepa2}. In this case the matrices $y$
and $w$ vanish and the constraint involves the eigenvalues of the $x$
matrix.

The second case where the constraint simplifies is for a single $z$
field, i.e. $i=\{1\}$, such that there is only one $f$ and $h$
function.  Then the matrices $x$,$y$ and $w$ become scalars 
\ba b &=& g_{_X} g^{_X}-3 \label{defs}
\\
x &=& h^{-1} (f -f_{_{\bar X}} g^{_{\bar X}}),
\nn \\
w &=& - g^{_X} g^{_{\bar X}} h^{-1} (h_{_{X \bar X}} - h_{_X} h^{-1}
h_{_{\bar X}}),
\nn \\
y &=& h^{-1} g^{_X}\[ f_{_X}-2 h_{_X}h^{-1}f-f_{_{\bar X}}
g_{_{X}}^{_{\bar X}} +\( f_{_{\bar X}} g_{_{XX}}^{_X} + h^{-1}h_{_X}
f_{_{\bar X}} -f_{_{X \bar X}} \) g^{_{\bar X}} \].  \nn \ea
For a canonically normalized $z$ field, $h =1, \, h_{_X} = h_{_{X\bar
    X}} =0$ which implies $w=0$.

For single field inflation, or if the matrices $x,w,y$ can be
diagonalized simultaneously, the eigenvalues of the mass matrix are
given by
\be 
m_\pm^2 =  \e^g \[ (1+b)+ |x|^2 +w \pm | (2+b)x +y| \].
\label{mass2}
\ee
The $z$ eigenstates remain stabilized as long as the smallest mass is
positive definite $m_-^2 > 0$.


\subsection{Separable K\"ahler function}
\label{s:adding}

The results in the previous section are a generalization of the work
\cite{Kepa1,Kepa2,Kepa3}, who considered a set-up with separable
K\"ahler functions:
\be
G(X,\bar X,z_i ,\bar z_i) = g(X,\bar X)+ \tilde g(z_i ,\bar z_i),
\label{Gsimple}
\ee
which is a special limit of the more general function \eref{G}. 
For the separable K\"ahler function above \eref{Gsimple} all mixed
derivatives of $G$, such as $G_{z z X}$, cancel. With this
simplification
\be
b = g_{_X} g^{_X} -3, \qquad 
x^{\bar i}_m = \tilde g^{\bar i k} \tilde g_{km}, \qquad
y^{\bar i}_j = w^i_j =0. \label{defx}
\ee
We now consider the case with only one $z$ field, which turns $x^{\bar
  i}_j$ into a scalar. As one can always diagonalize $x^{\bar i}_j$,
this simplification precisely gives the result along one of the
eigenvectors, and thus can be straightforwardly be generalized to
several $z$ fields. We recover the system studied in
\cite{Kepa1}\footnote{Our definition of $b$ is different from
  \cite{Kepa1}, which has $b \leftrightarrow b-3$.}:
\be
M^{z}_{z}|_{z_0} = \e^g[(b+1) + |x|^2], \qquad
M^{\bar z}_z |_{z_0}= \e^g(b+2)x,
\ee
which has eigenvalues
\be m_\pm^2 |_{z_0} = \e^g\[ 1+b +|x|^2 \pm |(2+b)x| \]
=\e^g\[ \( |x|\pm\frac{1}{2}|b+2|\)^2-\frac{b^2}{4}\].
\label{msep}
\ee
This result can also be obtained from the general expression for the
mass squared eigenvalues \eref{mass2}, taking the appropriate limit
$y^{\bar i}_j=w^i_j=0$. The function $b$ is bigger, equal or smaller
than zero for a dS, Minkowksi or AdS universe, respectively. Take $b
\geq 0$; in the opposite limit the masses $m_-^2$ and $m_+^2$ are
exchanged.  The smallest mass eigenstate is positive $m_-^2 >0$, i.e.,
the $z$-field is stabilized along the inflationary trajectory, for
$|x| <1$ or $| x|> (1+b)$. We will put this analysis in practice for
sgoldstino inflation in subsection \ref{s:hybrid} (hybrid inflation)
and \ref{s:small} (small field inflation).

Close to the instability bounds $|x| \lessapprox 1$ or $| x|
\gtrapprox (1+b)$ the spectator field $z$ is lighter than the
gravitino mass and/or the Hubble scale, and cannot be integrated out.
In a Minkowski vacuum after inflation either $b = 0$ or $b\to\infty$;
the latter case may occur in a supersymmetric vacuum with $W \to 0$.
For $b=0$, the masses reduce to $m_\pm^2 = m_{3/2} ^2\left(1 \pm
  |x|\right)^2$, with $m_{3/2}$ the gravitino mass.  For $|x|>1$, the
lightest scalars from the supersymmetric sector are heavier than the
gravitino. However, for $|x|<1$ the lightest of the two eigenstates is
lighter than the gravitino and cannot be neglected from a low--energy
description. This will play an important role later.  In the
supersymmetric vacuum with $b \to \infty$ we find $m_\pm^2 \approx
V_0(1\pm |x|) \to 0$, and the spectators are massless.  To avoid a
plethora of massless fields in the theory, one has to either break the
supersymmetry, or else go beyond the simple separable form of the
K\"ahler function \eref{Gsimple}.


\section{Single field sgoldstino inflation}

In this paper we focus on effectively single field inflation models,
for simplicity.  The inflaton, a real scalar, is identified with a
suitable linear combination of the real and imaginary parts of the
sgoldstino field; the orthogonal combination is to remain stabilized
at a local minimum of the potential during inflation. Single field
inflation can be divided into three classes: small field, large field
and hybrid inflation.  In the first two cases, if the model only
contains a single chiral superfield, the inflaton is automatically the
sgoldstino. If several fields are present, as is the case for hybrid
inflation, one has to be more careful, as the sgoldstino does not have
to coincide with the inflaton direction.

As is well known any sugra model of inflation has to address the
$\eta$-problem \cite{Dine:1995kz,Lyth:2004nx,Copeland:1994vg}: the
inflaton field needs to be protected from its natural tendency to
become heavy, and obtain a mass of the order of the Hubble scale. This
is just another manifestation of the hierarchy problem that plagues
all scalars, including the standard model Higgs field.  The problem is
easily spotted in the sugra context.  Expand the K\"ahler potential
around $X_0$, the inflaton field value during inflation, in $\delta X
= X-X_0$; this gives $K = K_0 + K_{_{X \bar X}}\big|_{0} \, |\delta
X|^2 + ... = K_0 + |\Phi|^2 + ...$, with $|\Phi|$ the canonically
normalized complex field. The scalar potential then gives
\be
V_F = \e^{|\Phi|^2}[V_0+ ...].
\label{eta_problem}
\ee
The $\eta$-parameter measures the curvature of the potential in units
of the Hubble parameter along the inflationary trajectory: $\eta=
V_{\phi\phi}/V$, with $\phi$ the canonically normalized real inflaton
field.  With the inflaton some linear combination of the real and
imaginary parts of $\Phi$, it is clear that the exponent in
\eref{eta_problem} contributes order unity: $\eta \approx 1 + ...$,
which spoils inflation.

The $\eta$-problem may be solved introducing symmetries which forbid
an inflaton mass, and thus keep the inflaton potential flat.  Such a
symmetry needs to be softly broken to provide a small slope for the
inflaton potential. It is far from trivial to assure that such a
breaking does not introduce the $\eta$-problem again.  Another
solution to the $\eta$-problem is to fine-tune parameters.  The order
one contribution coming from the exponent in \eref{eta_problem} may be
tuned against all other contributions (from the ellipses) to obtain a
total $\eta$-parameter that is small.

In the remainder of this section we will discuss large field, small
field and hybrid sgoldstino inflation, and how the $\eta$-problem may
or may not be addressed in each case.

\subsection{Large field inflation}

In models of large field inflation \cite{Linde:1983gd}, the inflaton field traverses
super-planckian distances in field space during inflation.  For a
potential dominated by a single monomial during inflation, $V \sim
\lambda \phi^n$, the slow roll parameters
\be
\eps = \frac12 \(\frac{V_\phi}{V}\)^2, \qquad \eta =
\frac{V_{\phi\phi}}{V},
\label{sr}
\ee
both scale as $\eta,\eps \sim 1/\phi^2$, and are automatically suppressed
for super-planckian field values. At first sight, no tuning of the
potential is needed. However, the problem is that for such large field
values {\it all} non-renormalizable operators are
unsuppressed. Therefore, an explicit UV completion of the model is
needed to determine whether inflation is possible.

Embedding large field inflation in sugra provides a better control
over the UV behavior of the theory.  Because of the $\eta$-problem
such an embedding is far from straightforward, as the potential
\eref{eta_problem} grows exponentially rather than
polynomial. Fine-tuning $\eta$ is not an option, as $\eta$ has to be
small along the whole inflationary trajectory, which spans
super-planckian distances in field space $\Delta \phi >1$.  This is in
contrast with small field inflation, discussed in subsection \ref{s:small}, where
the $\eta$-problem can be solved by tuning $\eta$ at a single point in
field space.  

Instead of fine-tuning, we can try to solve the $\eta$-problem by invoking a shift
symmetry \cite{kawasaki}. Consider a K\"ahler function $G = {\cal K}(X - \bar X)$, which is
symmetric under a shift $X \to X + c$ with $c$ a real constant.  Since
$G$ does not depend explicitly on $\phi \propto {\rm Re}(X)$, the
exponent in \eref{eta_problem} is independent of $\phi$ and there is
no $\eta$-problem.  In fact, the potential has an exactly flat
direction. Since we want the system to end up after inflation in a
Minkowski minimum, there is no other option than to set $V =0$ along
the flat direction, which is incompatible with having inflation.

In order to get a slope for the potential and obtain inflation, the
shift symmetry needs to be weakly broken. To assure the breaking does
not reintroduce exponential terms that ruin inflation, we add a
logarithmic term $G = {\cal K}(X-\bar X) + \ln|W(X)|^2$ with $W$ not
growing faster than power law. As we want to construct a potential that is polynomial, we forbid the linear terms in the K\"ahler potential ${\cal K}={\cal K}\((X-\bar X)^2\)$. Then
the potential along the inflationary trajectory is
\be
V_F|_{X = \bar X} = W_X G^{X\bar X} \bar W_{\bar X} - 3|W|^2\big|_{X = \bar X}.
\ee
The inverse metric $G^{X\bar X}|_{X =\bar X} =-1/{\cal K}''(0)$ is a constant
along the inflationary trajectory, as it is independent of $\phi$; it
just renormalizes the field and can be absorbed by going to
canonically normalized fields: $\phi^2 = -2{\cal K}''(0) |X|^2$.  If the
superpotential during inflation is dominated by a monomial term $W
\sim \lambda X^n$, we find
\be
V_F|_{X = \bar X} \propto  n^2 \phi^{2n-2} - 3\phi^{2n}
\label{VF_largefield}
\ee
which goes negative for large $\phi > n/\sqrt{3}$. For field values
$\phi = {\mathcal O}(10)$ as needed for large field inflation, the
field will run off to infinity and negative potential, rather than the
Minkowksi minimum at the origin. This does not give a viable inflation
model.  The instability occurs for every superpotential that does not
grow faster than power law, such that the shift symmetry is only
broken softly.  Faster growing superpotentials reintroduce the $\eta$
problem.

Although we did the analysis for a single field, this
straightforwardly generalizes to the multi-field case. If the inflaton
is the sgoldstino, it decouples from the other fields, and its
potential can be analysed independently and will always be of the form
\eref{VF_largefield}.  We conclude that large field sgoldstino
inflation in a sugra model does not work as it is plagued by an
instability in the scalar potential.

We note that it is certainly not impossible to have large field
inflation in sugra, only that it does not work with a single chiral
superfield.  Two-field models have been constructed that avoid the
instability \cite{rube,kawasaki}, employing a shift symmetry to
address the $\eta$-problem. However, in these models the inflaton is
{\it not} the sgoldstino (rather the sgoldstino is the orthogonal
field).

\subsection{Hybrid inflation}
\label{s:hybrid}

Hybrid inflation is a multi-field model of inflation which in addition
to the inflaton contains one or more so-called waterfall fields, which
serve to end inflation \cite{HI}. During inflation the waterfall fields are
stabilized in a local minimum, and inflation is effectively single
field. If the inflaton field drops below a critical value one of the
waterfall fields becomes tachyonic, and inflation ends with a phase
transition.

Standard F-term hybrid inflation \cite{HI_susy,HI_sugra} is an example
of sgoldstino inflation.  The K\"ahler function is of the separable
form \eref{Gsimple} discussed in section \ref{s:adding}.
\be
G = g(X,\bar X) + \tilde g(\chi_1, \bar\chi_1, \chi_2, \bar \chi_2),
\ee
with\footnote{To see that this setup is indeed of the general form
  \eref{G}, one can move a factor of $\ln|\mu^2|^2$ from $\tilde{g}$
  to $g$ and Taylor expand the remaining $\ln|\frac{\chi_1
    \chi_2}{\mu^2} - 1|^2$.}
\be
g = X \bar X + {k_s}(X \bar X)^2 +
\ln|\kappa X|^2 + ..., \qquad
\tilde g = \chi_1 \bar \chi_1
+\chi_2 \bar \chi_2 + \ln|\chi_1 \chi_2 - \mu^2|^2+...\nn
\label{hybridsetup}
\ee
The model has an R-symmetry, which uniquely fixes the superpotential
at the normalized level, and in particular it allows for a linear term
in $X$ but forbids the quadratic and cubic terms in $W$.  This kills
large contributions to the slow roll parameters, and allows for a flat
direction in the inflaton potential, which at tree level is only
lifted by higher order terms in the K\"ahler potential.

\begin{figure}[t!]
\begin{center}
\includegraphics[scale=0.75]{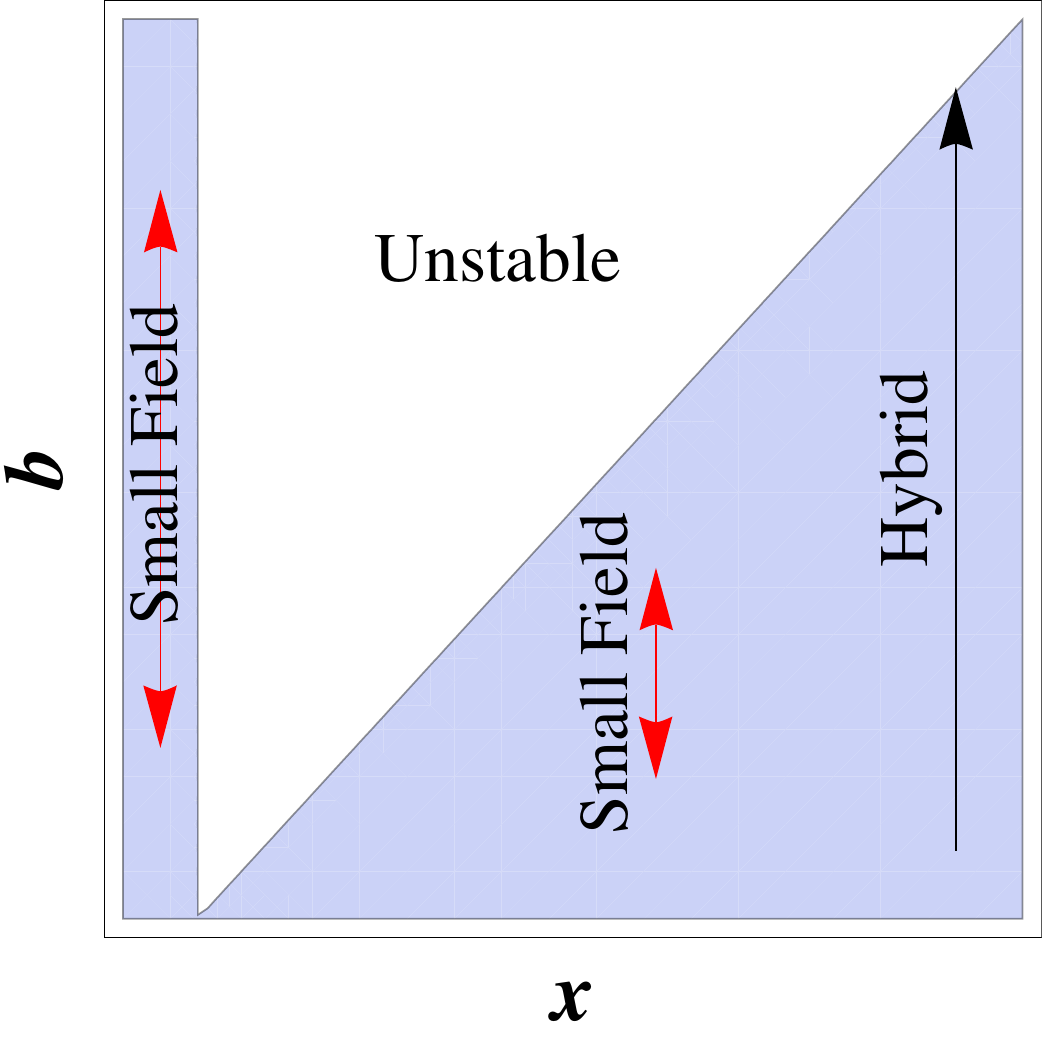}
\end{center}
\caption{\footnotesize (Figure adapted from \cite{Kepa1,Kepa3}.)
  Stability diagram for the separable case $G = g(X,\bar X) + \tilde
  g(z,\bar z)$. The variables on the axes $b,\, x$ are defined in
  \protect \eref{defx}, with $x$ one of the degenerate eigenvalues
  of the $x^{\bar i}_j$ matrix.  The masses of the spectator fields
  are positive in the shaded region, while the unstable region signals
  a tachyonic mode. The black arrow represents the inflationary
  trajectory for the proposed hybrid set-up, which ends when one of
  the spectator fields (the waterfall fields) becomes tachyonic. Also
  shown are possible inflationary trajectories for small field
  inflation (red arrows).}
\label{stability}
\end{figure}

The inflaton $\phi$ is identified with the real direction via the
decomposition $X = (\phi + i\theta)/\sqrt{2}$.  Inflation takes place
for large $\phi > \phi_c = \sqrt{2}\mu $, and all other fields
stabilized at zero field value.  The potential along the inflationary
trajectory is
\be
V = \kappa^2 \mu^4\( 1-  2 k_s \phi^2 +...\) + V_{1-{\rm
    loop}}.
\ee
The flatness of the potential is only lifted by higher order terms in
$K$, and by the one-loop Coleman-Weinberg potential $V_{1-{\rm loop}}$
\cite{Coleman:1973jx} .  The $\eta$-problem is solved via a moderate
fine-tuning of $k_s \lesssim 10^{-2}$. Moreover, for the 1-loop
contribution to be sufficiently small $\sqrt{\kappa}\mu$ should be of
the grand unified scale or smaller.  During inflation
$G_X=\frac{\sqrt{2}}{\phi}+\frac{\phi}{\sqrt{2}}+\frac{k_s
  \phi^3}{\sqrt{2}}$ and $G_{\chi_1} = G_{\chi_2} = 0$.  Hence $\phi$
is indeed the (real part of the) sgoldstino field.

The Minkowski minimum after inflation is at $X=0$, and $|\chi_1|=|
\chi_2| = \mu$.  In the minimum $G_X = G_{\chi_\pm} =0$ and susy is
restored.  There is no relation between inflation and low energy susy
breaking.  The sgoldstino during inflation is unrelated to the
sgoldstino today.

The masses of waterfall fields along the inflationary trajectory can
be found using the results of section \ref{s:adding}.  The mass
eigenstates are the linear combinations $\chi_\pm = (\chi_1 \pm
\chi_2)/\sqrt{2}$. Using these as a basis the matrix $x^{\tilde{i}}_m$
becomes diagonal during inflation. This shows that we can
restrict our attention to only one of the complex fields $\chi_\pm$,
the other field will give the same masses for its two real degrees of
freedom. Now we can directly compute the masses from \eref{msep}. The
stability region as a function of $b$ and $|x|$ is plotted in Fig
\ref{stability}.  The inflationary trajectory corresponds to a
vertical trajectory in the plot, going upwards as the field rolls
down.  When it irrevocably hits the instability region (i.e. when the
lower mass eigenvalue becomes negative), inflation ends.

We note that a similar stability analysis can be done for all models
of sgoldstino inflation.  Whereas hybrid inflation critically makes
use of the instability regions, for any non-hybrid scenario --- being
it small or large field inflation --- the inflationary trajectory
would have to stop before reaching the instability region.  This is
automatic for $|x|<1$, otherwise the stability conditions place an
upper bound on $b$ during inflation. We will return to this point
shortly when discussing small field inflation.

\subsection{Small field inflation}
\label{s:small}

Inflation in small field models \cite{{Linde:1981mu},Albrecht:1982wi}
takes place for sub-Planckian values of the inflaton field. This
allows for Taylor expanding the inflaton potential around its
Minkowski minimum. If one term in the polynomial expansion dominates
during inflation the slow roll parameters blow up: $\eps,\eta \sim
1/\phi^2$ in the small field limit, prohibiting inflation. As before,
$\phi$ is the canonically normalized inflaton field. The only way to
get around this conclusion is that several terms in the expansion
conspire together to nearly cancel, thus obtaining small slow roll
parameters.

This motivates to consider inflation near an extremum --- a maximum,
saddle point or inflection point --- of the potential.  This assures
that the first slow roll parameter $\eps$ vanishes. The
$\eta$-parameter can be made small by tuning the parameters in the
potential.  Since the inflaton field traverses only small,
sub-planckian distances in field space, tuning the curvature of the
potential at a single point, at the extremum, suffices.  This is in
contrast with large field inflation, where $\eta$ needs to be small
along the full, super-planckian inflationary trajectory. The tuning of
the parameters in the potential is typically of the 1-permille level,
dictated by the need to get $\eta \lesssim 10^{-2}$. Note that in a
sugra the $\eta$-parameter cannot be tuned for arbitrary K\"ahler
geometry \cite{Covi1,Covi2, BenDayan}.  In our example below we will assume an
(approximately) canonical K\"ahler potential, for which there are no
obstacles. Ref. \cite{BenDayan} considered modular inflation near a
maximum; we come back to this model at the end of this section.
 
Symmetries generically do not help in solving the $\eta$ problem in
the small field models. For example, a shift symmetry $K = K(X - \bar
X)$, so useful in large field models, does not do anything in the
small field regime. By Taylor expanding the K\"ahler potential and
performing a K\"ahler transformation, it becomes equivalent to a non
shift symmetric $K = K(X \bar X)$.  R-symmetries may help in providing
a flat potential, but the R-symmetry breaking, which is necessary to
obtain a Minkowski vacuum, also tends to spoil the flatness.  This is
what kills the model proposed in \cite{dine}, on which we will comment
in a bit more detail below.

We were able to construct a fine-tuned small field inflation model in
sugra containing only a single chiral field. In such a set-up the
inflaton is automatically the sgoldstino, and our example is an
existence proof for small field sgoldstino inflation.  Consider a
model with\footnote{ This ansatz \eref{KWsmall} is equivalent to $
G= \sum_{n=1}\alpha_n (X \bar X)^n + \log |\sum_{n=0}\lambda_n X^n|^2$.}
\be
K = \sum_n\alpha_n (X \bar X)^n, \qquad 
W = \sum_n \lambda_n X^n. \label{KWsmall}
\ee
We decompose the complex scalar $X= (\phi+i\theta)/{\sqrt{2}}$ with
$\phi$ the inflaton field. The model parameters $\lambda_n,\alpha_n$
can be tuned in such a way that the potential allows for inflation
near an inflection point which, without loss of generality, is located
at the origin $(\phi,\theta) = (0,0)$, and a Minkowski minimum at
finite field value $(\phi,\theta) = (\phi_0,0)$.  In particular, we
demand
\begin{itemize}
\item{Vanishing slope and curvature of the potential at the origin
 1) $V_\phi |_{(0,0)}=0$ and 2) $V_{\phi\phi}|_{(0,0)}=0$, to assure zero slow roll
  parameters $\eps = \eta =0$.  The condition on $\eta$ may be relaxed
  to $\eta \lesssim 10^{-2}$.}
\item{The height 3) $V|_{(0,0)} \equiv V_0$ of the potential at the origin
  is fixed by the COBE nor\-ma\-li\-za\-tion of the inflaton perturbations.}
\item{After inflation the inflaton settles in a local Minkowski minimum
    with 4) $V|_{(\phi_0,0)} =0$ and 5) $V_\phi|_{(\phi_0,0)} =0$.
    Moreover, the masses are positive definite 6) $m_i^2|_{(\phi_0,0)} >
    0$.}
\item{Along the whole trajectory, from the extremum to the minimum, the
  orthogonal field is stabilized 7) $V_\theta=V_{\phi\theta}=0$
  and 8) $m^2_{\theta} \gtrsim H^2$.}
\end{itemize}
We consider solutions with canonical kinetic terms, i.e. we set
$\alpha_1 =1$ and $\alpha_i =0$ for $i\neq 1$.  To satisfy conditions
1-5 we need at least five parameters and choose them accordingly. We
take all $\lambda_i$ real, and consider the first five in the
expansion.  Tuning is required to satisfy conditions (2) and (4) ---
the smallness of $\eta$ parameter and of the cosmological constant ---
in the usual sense that large contributions should nearly cancel.
Conditions 6-8 are then checked for consistency, but do not require
any new input.  Setting the minimum at $\phi_0 =1$ we find two
inflationary inflection point solutions\footnote{$\lambda_3 =0$ only
  vanishes for $\phi_0 =1$, but is non-zero for other positions of the
  minima.}
\be
\{\lambda_0,\lambda_1,\lambda_2,\lambda_3,\lambda_4\}=
\sqrt{\frac{V_0}{23}}\times
\{ 3,  -5\sqrt{2},3, 0,2\} ,
 \label{example1}
\ee
and
\ba
&&\{\lambda_0,\lambda_1,\lambda_2,\lambda_3,\lambda_4\}=
\frac{\sqrt{V_0}}{19\sqrt{73}}   \times
\label{example2} \\
&&
\bigg\{{3} \sqrt{39287-1464\sqrt{6}},
\sqrt{2\left(543551-19764\sqrt{6}\right)}, {3}
\sqrt{39287-1464\sqrt{6}},0,
-{2}
\sqrt{4943-1152\sqrt{6}} \bigg\},
\nn
\ea
and all other $\lambda_i$ are zero.  

\begin{figure}[t!]
\begin{center}
\includegraphics[scale=0.7]{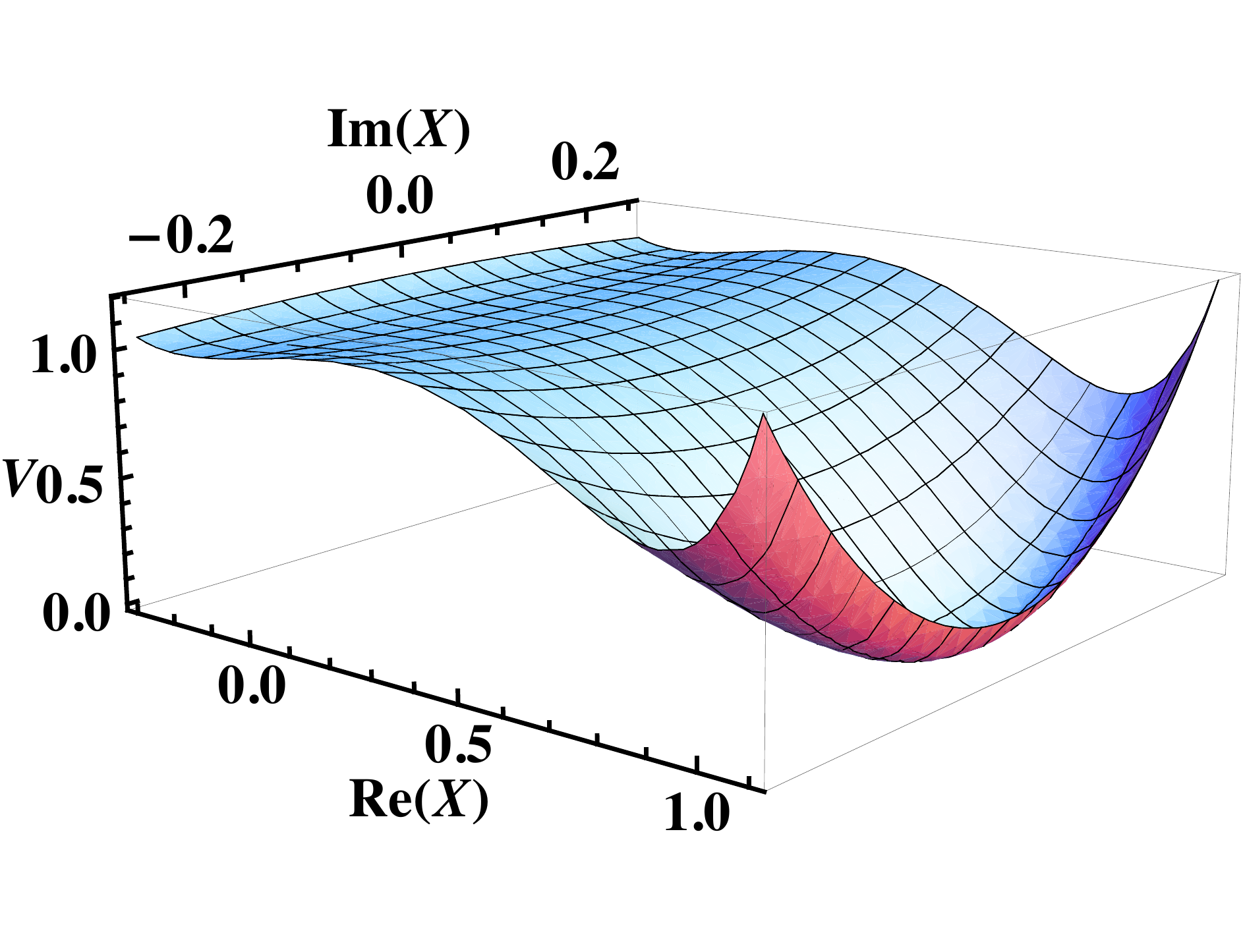}
\end{center}
\caption{\footnotesize Scalar potential for small field inflation
  corresponding to the first solution \protect \eref{example1}.}
\end{figure}

Both examples above correspond to inflection point inflation, rather
than to inflation near a maximum or saddle point.  This is
unfortunate, as for inflection point inflation the spectral index is
bounded to be $n_s \lesssim 0.92$, which is on the verge of being
ruled out. We review this argument in appendix \ref{s:inflec}.

The spectral index can be larger if the cubic term is absent or
unnaturally small, as is the case for inflation at a maximum rather
than an inflection point.  Then the correction to the spectral index
\eref{n_s} is set by the quartic term in the Taylor expansion around
the extremum, rather than by cubic term, with an upper bound $n_s
\lesssim 0.95$.  In our set-up this would require an extra tuning
condition $V_{\phi\phi\phi} \approx 0$; without it we always find a
saddle point.

\begin{figure}[t!]
\begin{center}
\includegraphics[scale=0.75]{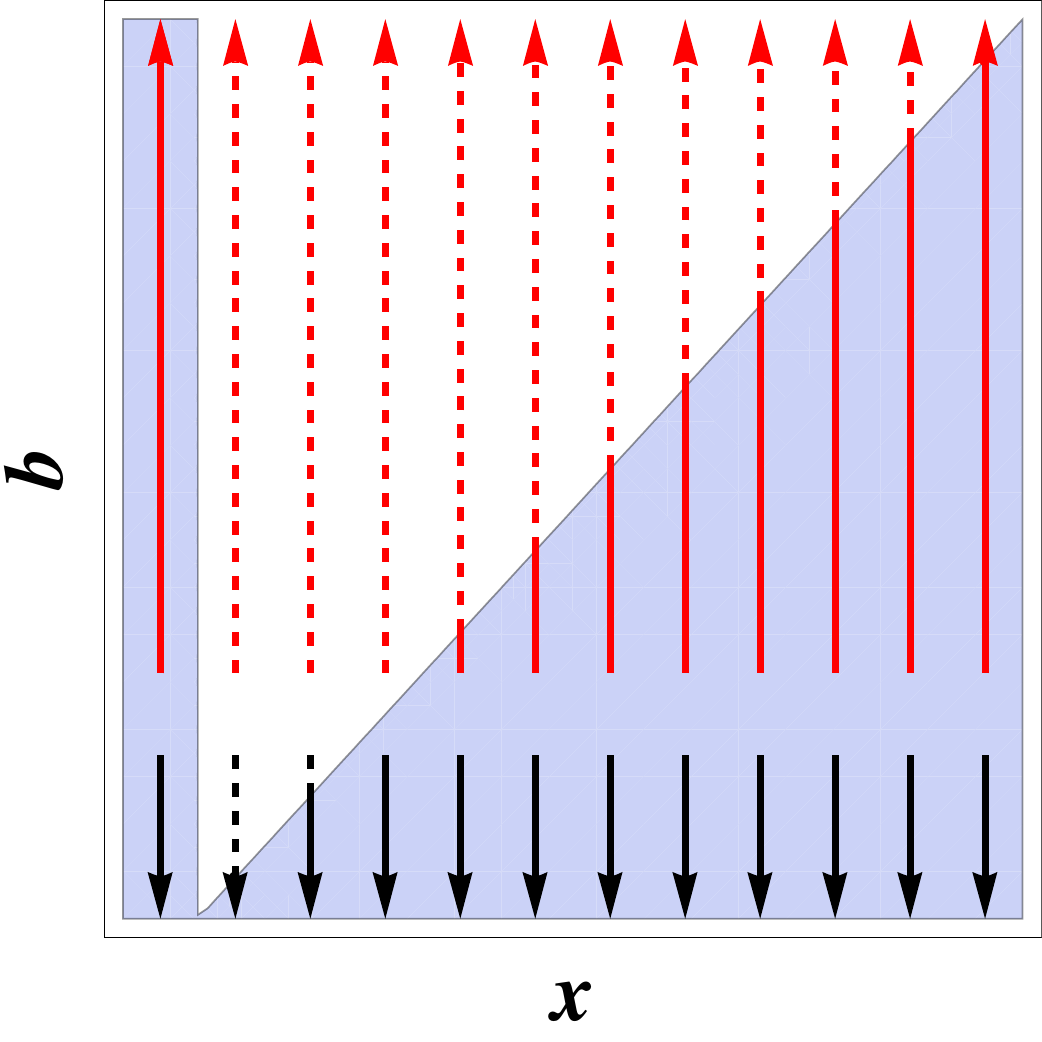}
\end{center}
\caption{\footnotesize Stability plot of the spectator $z$-fields for
  a separable K\"ahler function $G = g(X,\bar X)+\tilde g(z,\bar z)$.
  The trajectories for small field inflation are vertical lines, going
  upward (red) to infinity for solution \protect \eref{example1} which
  has a susy preserving vacuum, and downward (black) to zero for
  \protect \eref{example2} which has a susy breaking vacuum. Dashed
  lines indicate unstable trajectories. The position on the horizontal
  axis depends on the specifics of the spectator sector.  Solution
  \protect \eref{example1} always leads to an instability for $|x|
  >1$.}
\label{f:smallfield}
\end{figure}

The first solution above \eref{example1} has a supersymmetric
Minkowksi minimum. In this scenario the susy breaking observed today
is not related to the susy breaking during inflation. The second
solution \eref{example2}, however, does end in a susy breaking
minimum, and the gravitino mass today can be related to the
inflationary scale.  The gravitino mass is $m_{3/2} \sim 10^{-7}$, see
appendix \ref{s:inflec}.

There is a huge difference between the two solutions when combined
with other spectator fields.  The first solution has a susy preserving
vacuum in which $W \to 0$.  Although at this exact point our
description in terms of a K\"ahler function $G$ breaks down, we can
nevertheless describe the behavior of the potential as we approach this
singular limit.  We find that $b \propto V_0/W_0 \to \infty$, with $b$
defined in \eref{defs}. This implies that if we draw the stability
diagram for the simplified case of separable K\"ahler functions \eref{Gsimple},
see Fig.~\ref{f:smallfield}, this inflationary model corresponds to
vertical trajectories going upwards to infinity.  

The position on the horizontal axis given by $|x|$ depends on the
specifics of the spectator sector, but it is clear that for all
$|x|>1$ one of the fields becomes tachyonic as the inflaton approaches
its minimum, and the potential is unstable.  Hence, solution
\eref{example1} with a susy vacuum can only be combined with different
fields if this extra sector has $|x| <1$ (for several fields the
eigenvalues of the $|x|^2$ matrix should all be less than unity). This
puts enormous limitations on the spectator sector.  For $|x|<1$ the
masses of the spectator fields vanish in the vacuum, as discussed at
the end of section \eref{s:adding}.  However, in a subsequent susy
breaking phase transition they may pick up a soft mass term.

This disastrous conclusion may be avoided by going to the most generic
K\"ahler function for sgoldstino inflation \eref{G} rather than
sticking to the separable case \eref{Gsimple}; it is hard to make a
general prediction as in the $b \to \infty$ limit also the other
quantities $x,w,y$ in the mass matrix \eref{mass2} may blow up.

In contrast, solution \eref{example2} has a susy breaking vacuum, and
the parameter $b = V_0/W = 0$ vanishes in the minimum.  The inflaton
trajectory again corresponds to a vertical trajectory in the stability
diagram, but now going downwards.  Except for a small region near $|x|
=1$ there are no instabilities in the potential, and at least for the
separable K\"ahler function \eref{Gsimple} sgoldstino inflation can
straightforwardly be combined with a spectator sector.  In the region
$|x|>1$ the spectator fields are heavy in the vacuum and can be
integrated out to get a low energy EFT.  In the other limit $|x| < 1$
the spectator fields are of the same order as the gravitino mass (see
the discussion at the end of section \ref{s:adding}), and are
relatively light.

Ref.~\cite{BenDayan} constructed a single-field potential with a
maximum, rather than an inflection point, suitable for inflation.  As
remarked above, this set-up gives a spectral index in better agreement
with the WMAP data than our inflection point model. The flat maximum
was obtained by only allowing odd powers in the superpotential $W =
\sum \lambda_{2n+1} \Phi^{2n+1}$, and fine-tuning the lowest four
$\lambda_{2n+1}$ parameters.  In the absence of a symmetry that can
guarantee this form of the superpotential, this model is more
fine-tuned than the inflection point set-up, as it also requires
tuning the even parameters $\lambda_{2n}=0$; not only the
$\eta$-parameter is tuned, but also $V_{\phi\phi\phi}$ at the extremum
should vanish.  We further note that in this set-up $W \to 0$ at the
maximum, and thus $b\to \infty$.  As discussed above, this puts
very strong constraints on the spectator sector, and may make it harder to
embed the inflaton model in a larger parent theory.

\subsubsection{Recent proposals for small field sgoldstino inflation}

In the recent literature there have been claims for small field
sgoldstino inflation, with no or very little fine-tuning of the
parameters in the potential.  As argued in this paper, unless some
symmetry principle is invoked, this is not possible as the slow roll
parameters generically blow up in the small field limit.  Indeed we
find that these proposals do not work, although the devil is sometimes
in the details.

Refs. \cite{minimal1,minimal2} propose a model of sgoldstino inflation
in a single field set-up without tuning of parameters.  To address the
$\eta$ problem they add a logarithmic term to the K\"ahler potential
\ba
K &=& X \bar X +  a X \bar X (X + \bar X) + b (X \bar X)^2 + ... - 2
\ln(1 + X + \bar X),\nn\\
W &=& f X + f_n M.
\ea
However, in the small field regime the logarithm can simply be
expanded and does not alter the qualitative structure of the
potential.  It also does not enhance the symmetry.

Taking arbitrary parameters, except for the constraint that the
minimum at the origin is stable and has zero cosmological constant,
both the epsilon and eta-parameter exceed unity throughout the whole
field space $|X| <1$.  Slow roll inflation cannot happen.  In
\cite{minimal1} it is actually claimed that $\epsilon<1$, but what
they calculate is $\eps_\theta =g^{\theta \theta} (V_{\theta}/V)^2$,
where we again decomposed the field $X = (\phi+i\theta)/\sqrt{2}$ and
$g_{ij}$ is the metric in field space.  However, in a situation where
the potential falls much steeper in the $\phi$-direction than in the
$\theta$-direction, this is not the relevant slow roll
parameter. Instead, one should use the more general multi-field
generalization $\eps = g^{ij} V_i V_j/V^2$.

Ref. \cite{minimal2} shows inflationary
trajectories with a large number of efolds $N > 60$.  However, their
trajectories are calculated in the --- non-applicable --- slow roll
approximation. For all initial points in field space proposed in
\cite{minimal1, minimal2} we have solved the full two-dimensional
field equations and the slow-roll approximations to them. In all cases
the slow roll solutions wildly diverge from the full solutions, which
can only give inflation for less than an efold, confirming once more
that this setup does not provide a slow roll regime.

The only way to get inflation in the set-up of \cite{minimal1,
  minimal2}  is to tune parameters near an extremum, along the lines
of our example \eref{KWsmall}.

Ref. \cite{dine} proposes a model with an approximate R-symmetry:
\be
K = S \bar S + \alpha (S \bar S)^2, \qquad W = W_0 + \mu^2 S -
\frac{\lambda}{2(n+1)}S^{n+1}.
\ee
The R-symmetry is only broken by $W_0$ and the higher order term in
the superpotential.  In the absence of the constant $W_0$, this
assures that the potential is nearly flat near the origin, as there
is no quadratic and cubic term in the superpotential. The potential is only
lifted by the higher order quartic term in the K\"ahler (which must be
tuned $|\alpha| < 10^{-2}$), and the 1-loop Coleman-Weinberg correction
(which vanishes at the origin).

The set-up looks ideal for inflation.  However, the $n$ degenerate
minima of the potential are all anti-de Sitter. To get a Minkowksi
minimum after inflation, the constant $W_0$ has to be turned on.  And
although this is a small correction to the potential near the minimum,
it is the dominant correction to the inflationary plateau at the
origin, and gives rise to non-zero slow roll parameters $\eps$ and
$\eta$.  We find that the resulting potential is too steep to generate
60 e-folds of inflation (at most a single efold is
possible). Moreover, the tilt of the classical potential (not
including the one-loop contribution, which may change this) is such
that, unless there is some initial velocity to make it roll uphill,
the inflaton will not end in the minimum which is lifted to $V=0$, but
rather in one of the other AdS minima.

For concreteness, we can choose to uplift the AdS minimum at positive
values of $\phi$ to a Minkowksi minimum (with $X =
(\phi+i\theta)/\sqrt{2}$).  Moreover, just as \cite{dine}, we take the
parameters in the superpotential real, which simplifies the analysis.
The resulting potential will have a positive slope at the origin, as
argued above, which kills inflation at the origin.  However, there
will always be a maximum of the potential in between the origin and
the minimum.  Can we do inflation here?  Although the R-symmetry has
lost all of its power here (as it can only help to keep the potential
flat near the origin), this is still a possibility. However, although
the epsilon parameter vanishes at the maximum, the $\eta$ parameter
naturally exceeds unity.  Of course, $\eta$ can be tuned, but as
follows from our analysis in section \ref{s:small}, to satisfy all
constraints one needs at least five parameters.  The potential of
\cite{dine} has not enough freedom to do so.  Moreover, adding extra,
say, higher order terms, and trying to tune $\eta$, we find that the
maximum morphs into an inflection point (although we did by no means
an exhaustive study).  This is as expected, there is no reason, no
symmetry, which assures that when expanded around the extremum as in
\eref{V_inflec}, the cubic term should vanish.

\newpage
\section{Conclusions}

Inflationary models in supergravity, where the inflaton sits in a
complex scalar superfield, necessarily involve a multifield analysis.
Any extra fields present during inflation must be integrated out to
give an effective single-field slow-roll dynamics that is consistent
with the CMB. However, even very heavy fields can leave a detectable
imprint in the spectrum of primordial perturbations, in particular
through a reduction in the speed of sound of the adiabatic
perturbations. The correct effective field theory for the adiabatic
mode has a variable speed of sound that depends on the background
trajectory. A necessary condition to recover the standard single-field
slow roll description is that the trajectory should have no turns into
the heavy directions. In this case, the speed of sound is unity, equal
to the speed of light, and integrating out the extra fields gives the
same effective action as truncating the heavy fields at their
adiabatic minima.

In supersymmetric models there is an extra complication. One
has to integrate out whole supermultiplets in order to obtain an
effective supergravity description for the remaining superfields. This
is only possible if the superfields that are being integrated out are
in configurations that do not contribute to susy breaking.

Sgoldstino inflation naturally implements these two conditions. The
full inflationary dynamics is confined to the sgoldstino plane.
Putting the scalar components of all other superfields at their minima
is a consistent truncation of the parent theory. This makes
sgoldstino inflationary models extremely attractive, because of their
simplicity and robustness.

We have analysed sgoldstino inflation scenarios exploiting the fact
that the K\"ahler in\-va\-riant function $G = K + \log |W|^2$ has a
relatively simple form \eref{G} which allows some aspects to be
analysed in a model--independent way.  We derived a necessary and
sufficient condition on the K\"ahler function (\ref{mass2}) for the
stability of the susy-preserving sector, the spectator fields that are
integrated out. Figure \ref{stability} shows the constraint for a
separable K\"ahler function, in particular for hybrid F-term inflation
(which is a well studied case of sgoldstino inflation).

In the case of small field sgoldstino inflation we were able to
provide some viable fine-tuned examples around inflection points. The
spectral index is rather low, on the verge of being ruled out by the
CMB data. A higher spectral index would be possible with additional
fine--tuning.  Rather surprisingly, the inflationary model can only be
straightforwardly combined with a spectator sector if the minimum
after inflation breaks susy.  In our inflation example with a susy
preserving Minkowski vacuum the spectator sector is very constrained
by the condition that there should be no tachyonic modes in the
system.  This is illustrated in Figure \ref{f:smallfield}. These
constraints would also affect the hilltop inflation examples in
\cite{BenDayan}.


One of the motivations for this study was the interesting suggestion,
put forward in \cite{minimal1}, that a relatively simple supergravity
model with a single chiral sgoldstino superfield could account for
both inflation and susy breaking in the vacuum. Contrary to claims in
\cite{minimal1,minimal2}, our conclusion is that this minimal scenario
is very tightly constrained and requires the usual level of
fine--tuning that is expected on general grounds.  Another interesting
model was proposed in \cite{dine}, in which the flatness of the
inflationary plateau follows from an R-symmetry. However we find that the
R-symmetry breaking needed to obtain a Minkowski vacuum introduces an
unacceptable tilt in the potential, and prevents inflation.  It is
possible that variations of this model may still work with some extra
fine--tuning.

\section{Acknowledgments}

We are grateful to Luis Alvarez-Gaum\'e, Michael Dine, C\'esar
G\'omez, Ra\'ul Jim\'enez and Lawrence Pack for discussions
of their models, to Jan-Willem van Holten for collaboration at the
early stages of this work, and to Lofti Boubekeur and Cristiano Germani for
discussions. PO is supported by F.O.M. (Dutch
organization for Fundamental Research in Matter).  AA
acknowledges support by Basque Government grant IT559-10, Spanish
Ministry of Science and Technology grant FPA2009-10612, the
Consolider-ingenio programme CDS2007-00042. SM and MP are supported by
NWO Vidi-grant (680-47-229).

\begin{appendix}
\section{Small spectral index for inflection point inflation} \label{s:inflec}

In this appendix we derive the spectral index and power spectrum for
inflection point inflation, following the work of Refs. \cite{brax,
  Linde:2007jn}. To a very good approximation the inflationary
observables only depend on the $\eta$-parameter at the extremum and on
the number of efolds.  

Expanding the potential around the inflection point gives:
\be
V = V_0(1+1/2 \eta_0 \phi^2 +C_3 \phi^3 + C_4 \phi^4 +...),
\label{V_inflec}
\ee
with $\eta,C_3 <0$ so that the field rolls towards the minimum at
positive $\phi$ values.  Inflation ends when the $C_3$ term becomes
important, and $\eps \approx 1$, which occurs for field values
$\phi_f^2 \sim \sqrt{2} /(3|C_3|) $.  We can calculate the number of
efolds
\be
N \approx \int_{\phi_f}^{\phi_N} \frac{V}{V'}
= \frac1{\eta} \log\[ \frac{\phi}{3C_3 \phi+\eta} \]^{\phi_N}_{\phi_f},
\ee
where we used $V \approx V_0$ above. The above expression can be
inverted to obtain the value of the inflaton field $N$ efolds before
the end of inflation $\phi_N$:
\be
\phi_N = \frac{\e^{N \eta_0} \eta_0/C3}{-3 (\e^{N \eta_0}-1)
  -\eta_0/(\phi_f C_3)} \approx \frac{\e^{N \eta_0} \eta_0}{-3 C_3 (\e^{N \eta_0}-1)},
 \ee
 where in the second step we used $\eta_0/(\phi_f |C_3|) \ll 1$.  This
 is a good approximation as $\eta_0 \ll 1$ is fine-tuned, whereas
 $C_3$, and thus $\phi_f$, is naturally of order one\footnote{To be
   precise, $C_3 = {\mathcal O}(1)$ for $\phi_0 \sim 1$.  For minima
   at smaller field values generically $C_3$ increases, as a sharper
   turnover of the potential is needed.  We do not find valid
   solutions for minima for $\phi_0 \gg 1$ much larger, as then other
   local minima at smaller field values appear.}.  Note that in this
 limit, the number of efolds is independent of the end of inflation,
 as $\phi_f$ has dropped out of the equation.  As a result the
 inflationary observables are insensitive to the precise coefficients of the
 higher order terms in \eref{V_inflec}. The spectral index is
\be 
n_s \approx 1 + 2 \eta \approx 1 + 2 \eta_0 +12 C_3 \phi_N \approx 1-2\eta_0 
\frac{(\e^{\eta_0 N}+1)}{(\e^{\eta_0 N}-1)},
\label{n_s}
\ee
where we used that $\eps \ll \eta$. For $N < 50-60$ one finds $n_s <
0.92-0.93$ for the whole range of $|\eta_0| \lesssim 10^{-2}$. The
power spectrum is
\be
P_\zeta = \frac{V}{150\pi^2 \eps} = \frac{3C_3^2 \e^{-4N\eta_0} (\e^{N
    \eta_0}-1)^4 V_0}{25\pi^2 \eta_0^4}
\ee
with $P_\zeta = 4 \times 10^{-10}$ measured by WMAP.

For the first example \eref{example1} in the text $\eta_0 = 0$ and
$C_3=-2.39$.  For $\eta_0 = 0$, the expressions simplify to 
\be
 n_s - 1=-\frac{4}{N}, \quad 
P_\zeta = \frac{3 C_3^2 N^4
V_0}{25\pi^2}, \quad
( {\rm for} \; \eta_0 = 0).
\ee
Choosing $N=50$ this gives $n_s = 0.92$ and $V_0 = 9 \times 10^{-16}$.
The second example \eref{example2} has $C_3 =-3.69$, and gives the
same spectral index and similar $V_0 = 4 \times 10^{-16}$.  The
gravitino mass today is related to the inflationary scale via $m_{3/2}
= \e^{K/2} W|_{\rm min} \sim 10^2 \sqrt{V_0} \sim 10^{-7}$, far above
the electroweak scale.

\end{appendix}


\end{document}